# Gender response to Einsteinian physics interventions in School

Tejinder Kaur[1], David Blair[1], Rahul Kumar Choudhary[1], Yohanes Sudarmo Dua[1], Alexander Foppoli[1], and Marjan Zadnik[1]

[1]*University of Western Australia, Crawley, WA, Australia, 6009.*
[2]*Curtin University, Bentley, WA, Australia, 6102.*

There is growing interest in the introduction of Einsteinian concepts of space, time, light and gravity across the entire school curriculum. We have developed an educational programme named 'Einstein-First', which focuses on teaching Einsteinian concepts by using simple models and analogies. To test the effectiveness of these models and analogies in terms of student attitudes to physics and ability to understand the concepts, various short and long interventions were conducted. These interventions were run with Years 6 to 10 academically talented and average IQ students. In all cases, we observe significant levels of conceptual understanding and improvement in student attitudes, although the magnitude of the improvement depends on age group and programme duration. This paper reports an unexpected outcome in regard to gender effects. We have compared male and female outcomes. In most cases, independent of age group, academic stream and culture (including one intervention in Indonesia), we find that female students enter our programmes with substantially lower attitude scores than males, while upon the completion of the programme, their attitudes are comparable to the boys. This provides a compelling case for widespread implementation of Einsteinian conceptual learning across the school curriculum. We discuss possible reasons for this effect.

## I. Introduction

In the last few decades, there is widespread concern about declining interest in STEM subjects, and gender equity. The issue of gender equity in science has been recognised for many years. Twenty years ago, Federico Mayor, who was a director-general of UNESCO made the statement, "On a worldwide scale, science... is still a man's business. This situation is no longer acceptable" [1], highlighting that this is still a significant issue. Ekine suggests that biases against girls "are manifested in science curricula, instruction and assessment" [2]. The European Institute for Gender equity emphasises the strong economic arguments for achieving gender equity in STEM education [3]. In 2014, Unterhalter et al reviewed the kind of interventions that lead to improved education outcomes [4].

Researchers around the world are investigating the underrepresentation of females in the STEM fields. The claim that female aptitude for mathematics and science is intrinsically lower than males [5], [6] is contradicted by recent results of the National Assessment of Educational Progress [7] and the Program for International Student Assessment (PISA) [8]. The results show that the gap between the achievement of male and female students in mathematics and science no longer exists. However, there is an attitude issue. It is suggested that there is a need to encourage them to gain a positive attitude towards STEM subjects [9], [10], [11]. In certain studies, it has been shown that female students have less positive attitude towards science than male students [1], [12]. Numerous studies have shown that high school is the time in which young female students' interest towards STEM subjects can be cultivated [13], [14], [15], [16], [17], [18], [19].

To create interest among female students towards physics, the development of female-friendly science is a commonly explored theme among researchers [15], [20], [21]. Research shows that the female-friendly strategies or connecting with everyday applications to teach science has a positive influence on females' perception of science [22], [23], [24]. Interactive discussions and experiences in physics classes have positive impact on male and female achievement in science [25], [26]. Dare and Roehrig reported, that Year 6 female students identified hands-on activities, interactive learning, and participation in experiments as the primary reasons in their enjoyment of science in schools [27]. Female students also identify that active participation or learning-by-doing helps them to remember things for longer rather than by just reading books. Students also mentioned that learning with models and analogies and working in groups give the opportunity for every student to participate. On the other hand, they reported that several boys preferred to work on their own instead of following instructions given by the teacher or working in a group [31].

Most of the research discussed above was undertaken in the context of the conventional Newtonian physics-based middle school curriculum, and investigated ways that the teaching could lead to improved female attitudes.

The focus of this research is completely different in that the gender outcomes were not derived from a deliberate strategy. The 'Einstein-First' project set out to investigate the possibility of introducing *Einsteinian* physics (a combination of special, and general relativity and quantum physics) as foundational concepts in school education [28]. The historical and useful Newtonian approximations are introduced later.

The Einstein-First project is founded on the idea that every child should learn humanity's best understanding of space, time, matter, light and gravity would seem self-evident. But this view is often countered by the view that modern concepts of physics are too difficult for children to learn. From the 1920s onwards, physicists and science writers [29], Einstein, Feynman and many others attempted to introduce the concepts of Einstein's General Theory of Relativity and Quantum physics to the general public. Many Newtonian concepts such as the absolute nature of space and time are an implicit part of primary school education.

Today, the concepts that light is a wave and that gravity is a force field emanating out planets are commonly taught. These concepts contradict Einsteinian concepts. While general relativity and quantum physics are beginning to be introduced in senior high schools, the learning of these topics are burdened by students' prior learning of classical concepts. In 1985 Feynman stated, "*I want to emphasise that light comes in this form – particles. It is very important to know that light behaves like*

*particles, especially for those of you who have gone to school, where you were probably told something about light behaving like waves*"[30]. This statement is still true in schools today.

The challenge of introducing Einsteinian concepts at an early age forced us to adopt and approach which, retrospectively, we recognise as being consistent with the female friendly approaches discussed above. However the necessity was not for reasons of being female friendly but for making Einsteinian physics generally accessible.

The Einstein-First project developed an approach consisting of activity-based learning using models and analogies, in an attempt to overcome the issue of difficulty [31], [32]. We developed activities that introduce students to fundamental concepts such as the shape of space, the nature of spacetime, the photon description of light, and the origin of gravity. Our motivation is to open students' minds to the concepts that underpin modern physics such as the relativity of space and time, and how quantum scattering underlies almost everything we observe around us. Even at much younger ages, we want students to be open to the idea that mathematics is more than arithmetic (for example it can include vector arithmetic which we introduce as the addition of arrows), and that geometry is more than Euclidean geometry, (indeed Euclidean geometry is merely a good approximation: students learn this by doing experimental geometry on woks).

We emphasise the relevance of Einsteinian science to our lives, which includes the technologies on which we rely, such as smart phones, and materials that we love such as gold, (created in the gravitational-wave driven coalescence of neutron stars and coloured by the relativistic motion of its electrons). We also emphasise to students that they are learning a story that has taken centuries to uncover; a story that is still incomplete, with more to be discovered in the future. We try to give students a sense of anticipation, that they are getting the first taste of concepts, which will be elaborated, in their future years of education. Our goal is to create a seamless progression of learning that begins with a valid, though simplified, representation of our best understanding of reality, that leads into more sophisticated and eventually, mathematical representations, as well as a clear understanding of how to revert to approximations, such as the approximation of Euclidean geometry which is almost always valid on Earth. This approach should ensure that future university teachers will not need to say, "forget what you learned at school" but rather to be able to build on concepts already introduced at school.

Unlike many of the previous studies, our research programmes and interventions were not designed to have a positive gender equalising effect. They were designed to ask the question: Is it feasible to introduce an Einsteinian approach across the school curriculum. While this yielded results that were not surprising to the research team – students' show excellent understanding, retention and positive attitudes to the programme - the gender effect was serendipitous. We show in this paper that the introduction of fundamental concepts, presented to mixed classes without any specific attention to gender yields surprisingly large gender equalising effects.

This paper focuses on the observed gender effect found in separate interventions with different age groups, different durations, different people delivering the programme, and even for a programme delivered in Bahasa Indonesian on the island of Flores.

In this paper, we confine our analysis to quantitative measures of the gender differences in student attitudes and knowledge before and after an intervention. First, we will describe the research methodology and summarise the interventions we have undertaken. Then, we will present a summary of the gender results and finally we discuss the implications of this work in relation to the K-12 curriculum, and make recommendations for implementation of an Einsteinian school curriculum.

## II. Methodology

### A. Participants and description of various interventions
In this study, 233 students (including academically-talented and average students) participated in seven different programmes and interventions. The age group of participating students was 11-16 years. Among these seven, four were one-day in duration and two were 10-weeks and one was three weeks long. The 10-week programme was run at Shenton College with academically-talented Year 9 students in 2013 and 2014, while the one-day programme was run with academically-talented Years 7, 8, 9 and 10 from Mount Lawley High School. These students were brought to the Australian International Gravitational Observatory (AIGO) research facility located near Gingin, north of Perth in Western Australia. The three-week programme was run in a high school in Indonesia. The brief description of every intervention is given below.

Table I. Description of every intervention of Einstein-First project.

| Interventions | School name | Year group | Number of students | Duration |
|---|---|---|---|---|
| 1 | Mount Lawley | Year 7 | 24 (13 males and 11 females) | One day |
| 2 | Mount Lawley | Year 8 | 16 (12 males and 4 females) | One day |
| 3 | Shenton College | Year 9 | 45 (24 males and 21 females) | 10 weeks |
| 4 | Shenton College | Year 9 | 57 (33 males and 24 females) | 10 weeks |
| 5 | Mount Lawley | Year 9 | 30 (14 males and 16 females) | One day |
| 6 | Mount Lawley | Year 10 | 30 (11 males and 19 females) | One day |
| 7 | Frateran Maumere Senior High School, Flores Island, Indonesia | Year 10 | 31 (18 males and 13 females) | 3 weeks |

### B. Nature of the programmes
All the programmes were designed by the authors of the study to introduce fundamental concepts of general relativity and quantum physics and were conducted during school hours. Number of lessons was chosen according to the duration of each programme. Every programme utilized both PowerPoint presentations and interactive hands-on activities. Every lesson was 45 minutes and it was structured as a) first 15 minutes for the presentation, b) next 15 minutes for the related activity, c) and the last 15 minutes for discussions or worksheets. The presentations introduced the concepts to the whole class in an interactive lecture environment where discussion was encouraged. The presentations were designed to explain Einsteinian Physics concepts visually and as a result included many pictures, videos, animations, and very few words. Activities reinforced concepts through active participation. Each activity was designed to capture an aspect of the learning concept and demonstrate it visually in a familiar setting. Thus reinforcing the otherwise 'abstract' concept.

### C. Data collection
This study is primarily interested in two criteria: the students' ability to understand the basic concepts of Einsteinian physics and their attitude towards Einsteinian physics. To measure the effectiveness of the programme, two tests were designed by the authors according to the procedure mentioned in section D. As there is a minimal or negligible literature exits in teaching and learning of Einsteinian physics, hence all the questions were created by the researchers.

A "conceptual pre–questionnaire" (see appendix) was designed to assess students' prior conceptual knowledge of Einsteinian physics. The pre-test had questions of 2 different types: open ended questions, two-tier questions (Yes/No with justification). This test was given at the beginning of the programme and 15 minutes were allowed to complete it.

The "conceptual post-questionnaire" was designed to assess students' conceptual knowledge of Einsteinian physics following completion of the programme. It had identical questions to the pre-test. This test was given at the end of the programme under identical conditions to the pre-test.

To measure the effectiveness of the programme on students' conceptual understanding, pre and post-tests were analysed and compared.

An "attitudinal pre-questionnaire" (see appendix) was designed to assess students' attitude towards physics. This test was based on the Likert scale items.

The "attitudinal post-questionnaire" was designed to assess students' attitude towards physics following completion of the programmes. It had identical questions to the pre-questionnaire. This questionnaire was given at the end of the every programme under identical conditions to the pre-test.

## D. Validity

The degree to which a proposed idea is accurately evaluated through test scores is known as validity. In order to ascertain the validity of both the knowledge and attitudinal questionnaires, the following questions were raised:

1. Do the questions encompass every topic we wish to teach the students and have these topics been addressed in the literature?
2. Are the students able to interpret the questions as they are meant to be understood?
3. Do educational experts agree that the questions are appropriate?

The following describes the extensive review process used to ensure the validity.

### 1. *Content and Literature validation*

In order to assess the students' understanding and attitude towards Einsteinian physics concepts, we used the topics we covered in the Einstein-first programme as the basis in designing the conceptual and attitudinal questionnaires, where only content-related elements were tested. Section I clearly indicated that hardly any research had been done into the teaching of basic Einsteinian physics theories. Therefore, several of the conceptual questions we asked had never been reported. Only a third of the questions were in current literature and thus were already validated questions. The rest of the questions, although innovative, were necessary and tailor-made for the programme.

### *2. Student interpretation validation*

Plain and concise words were needed in order for students to completely understand the questions asked. Several of the questions were ambiguous (i.e. 'What is light?' and 'Does space have a shape?') but these were explained so that students fathomed that the 'light' in this context was not the opposite of 'heavy' and that space referred to room or gap instead of the outer 'space'. Other questions were straightforward and directly related to the matters discussed in the programme.

### *3. Expert validation*

Experts on the subject including experienced physicists and educators reviewed the drafted questions, where each one was further deliberated on and refined. A database of existing physics questions was referred to while reviewing the conceptual questions. Questions were then reviewed and redrafted once more before being finalised for the study.

## D. Data analysis

### a) Conceptual understanding

Improvement in students' conceptual understanding as a result of the programme was evaluated by analysing the quality of conceptual pre and post-test answers.

The pre/post-test questions were assigned scores according to the different types of questions. Simple questions = 1 mark, two-tier questions = 2 marks. The students achieved marks according to their performance. In simple questions, they got 1 mark if they responded correctly. Half mark was given to the partial correct and no mark was given to the incorrect answers. In Two-tier questions, students had to justify their chosen answer. If a student chose right/wrong option and gave an incorrect/correct explanation, then 1 mark assigned. If students chose the right option and explained it correctly, then 2 marks were given to them. The two tests were assessed in the same way since their questions were identical.

By comparing the students' pre and post-test sum scores, we have a measure of the improvement in their conceptual understanding as a result of the programme. Furthermore, the t-test was performed to find any statistically significant difference in scores.

### b) Attitudes

For attitudinal questionnaires, students' responses for Strongly Agree and Agree as well as Strongly Disagree and Disagree were combined and converted into percentages, separately for males and females.

All the presenters used the same method to analyse the data. To calculate mean, standard deviation and t-tests, analyses were performed using Excel or SPSS.

## III. Results

This section reports research results obtained from different interventions. First, we present the results of students' conceptual understanding followed by the gender differences in conceptual understanding of Einsteinian physics concepts. In the next section, the results of gender differences in attitudes towards science are presented.

## A. Students results in conceptual learning

The males' and females' learning of Einsteinian physics concepts were compared according to their year group. In every programme, the males' initial knowledge of Einsteinian physics concepts was slightly higher than the females'. However, we found that in most of the interventions, the female students' improvement factor was higher than their male counterparts'.

In this paper, we present results obtained from three different interventions, two were conducted in Australia in 2013 and 2014 with Year 9 students (interventions 3 and 4) and the third one was conducted in Indonesia with Year 10 students (intervention 7). Intervention 4 is a refined version of intervention 3. In this intervention, the presenter was more confident than in the previous programme and the lesson plans were improved after gaining experience and feedback from intervention 3. Intervention 7 is a replication of intervention 4. The same knowledge and attitudinal questionnaires were used to collect the data. The methodology and lesson plans were also same. That is most likely the reason we obtained a similar trend of scores in intervention 7. The knowledge results of interventions 3 and 4 were extracted from another published paper [33]. Figures 1a and 1b below present results obtained from interventions 3, 4 and 7.

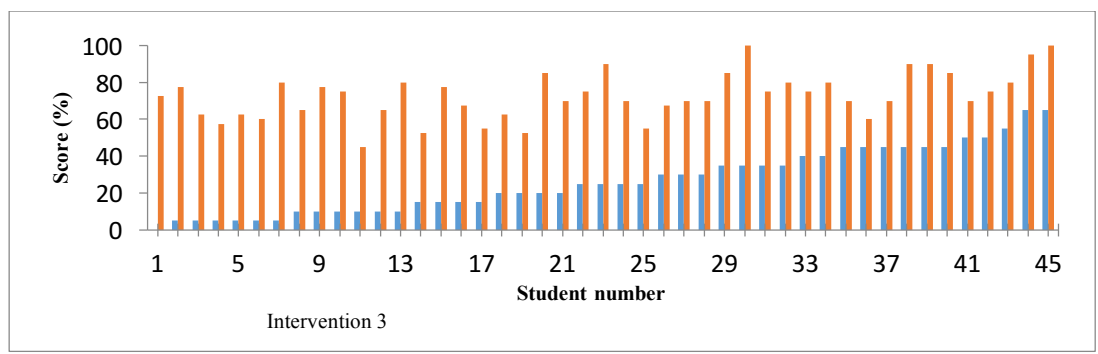

FIG. 1a. Conceptual understanding results obtained from Year 9 students in Australia i.e. Intervention 3.

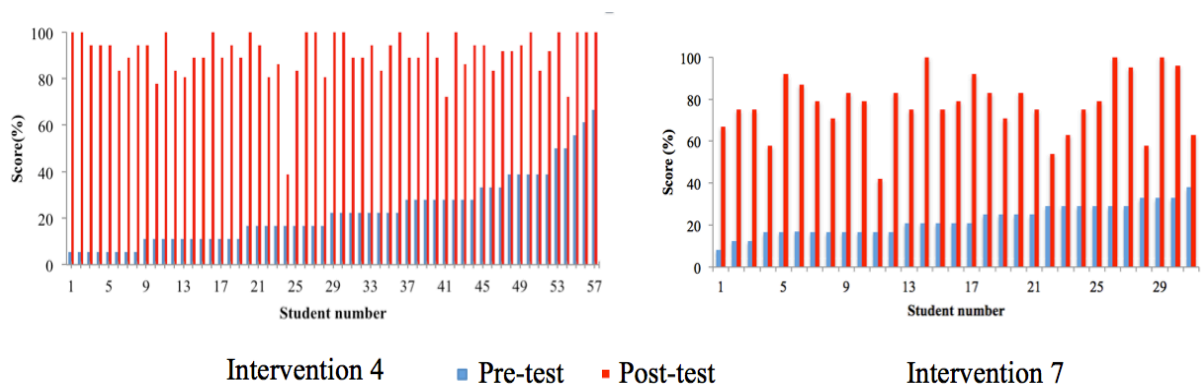

FIG. 1b. Conceptual understanding test results obtained from Year 9 students in Australia (intervention 4) and Year 10 students in Indonesia (intervention 7). The data in the figure clearly shows that students' pre-programme knowledge in Einsteinian concepts was low but the post-knowledge test results show a notable improvement.

It is clearly shown in figures 1a and 1b that the students' prior knowledge in Einsteinian physics was quite low in both countries. Around 56% of students from intervention 3, 77% of students from intervention 4 and 87% of students from intervention 7 achieved less than 30% in the pre-test. There were only a few students from all the interventions who achieved more than 50% in the pre-test. We asked students who managed to answer a few questions correctly in the pre-test, where they acquired their information of these modern concepts. Those students responded that they had watched a few science programmes and documentaries on TV.

After the programme, 31% of students in intervention 3 and 88% of students in intervention 4 achieved scores above 80% compared to 52% in intervention 7. In all the three interventions, the students who achieved low scores in the pre-tests improved maximally in their conceptual post-tests. No students showed any decrease in the scores after the programme.

The paired sample t-test was calculated for both interventions. For intervention 3, it was calculated as $t(44) = 20.1$, $p < .05$, for intervention 4, it was calculated as $t(56) = 30.3$, $p < .05$, while for intervention 7, it was $t(30) = 19.98$, $p < .05$, indicated that conceptual understanding improved after the Einsteinian physics programme. In all the three interventions, students' conceptual understanding towards Einsteinian physics improved significantly.

## B. Students' conceptual results according to the gender

The figure given below compares students' results in conceptual understanding according to their gender.

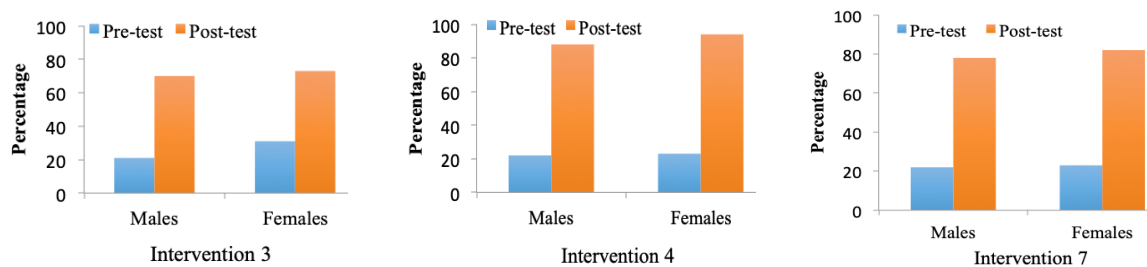

FIG. 2. Males and females' average scores in the knowledge pre-tests and post-tests. It is shown in the figure that females improved slightly more than their counterparts in all the interventions.

As shown in figure 2, both genders had very low average scores in the conceptual pre-tests. In all three interventions, the average score for both genders was around 23%. In both Australian interventions 3 and 4, the males' average score in the post-test was 70% and 88% while 78% was observed in intervention 7. On the other hand, the female students' average score in the post-test was noted as 73% in intervention 3, 94% in intervention 4 and 82% in intervention 7.

Overall, after every intervention, female students' knowledge improved slightly more than the males'.

## C. Gender differences in attitude towards science according to long intervention

First we present students' attitudinal results towards physics from long interventions. This section also presents the students' results towards learning modern ideas through activities followed by the brief explanation of attitudinal results obtained from short interventions.

# 1. Students' interest in physics

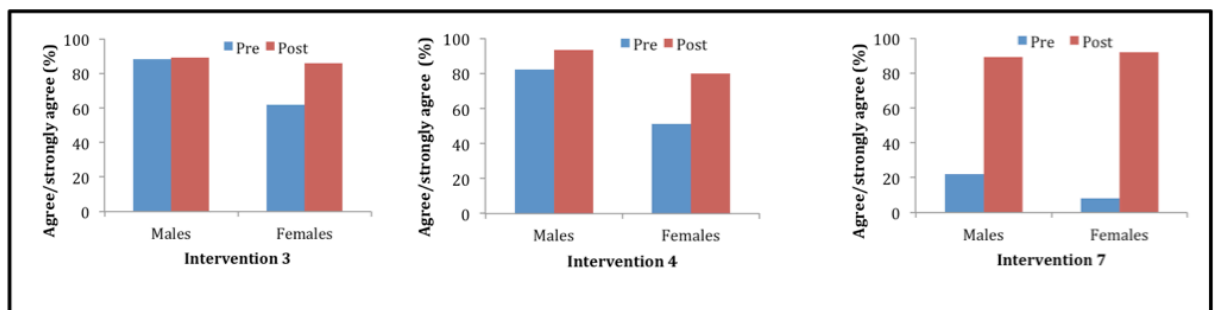

FIG. 3. This figure represents male and female students' interest in physics. The results were obtained from three interventions i.e. Intervention 3, intervention 4 and intervention 7.

Students' interest in physics was determined by asking the question "I think physics is an interesting subject". Figure 3 represents the results obtained from three different interventions. As shown in the figure, male students' interest in interventions 3 and 4 was high in the beginning compared to their female counterparts. In the pre-programme attitudinal test, the percentage of students who agree/strongly agree with the statement that physics is an interesting subject was 88% in intervention 3 and 82% in intervention 4. The percentages of female students who agree/strongly agree with this statement were 62% and 51% in interventions 3 and 4 respectively. These numbers dropped in intervention 7 where merely 22% of males and 8% of females found physics as an interesting subject.

After the 10-week programme on Einsteinian physics concepts, female students' attitude changed significantly. In intervention 3, females' interest was almost the same as the male students' as in intervention 4, where females' score matched the males' initial scores. In intervention 7, female students gave very positive response towards their interest in physics. After attending a three-week programme on modern physics concepts, 92% of females found physics as an interesting subject. There was also significant improvement observed in male students, where 89% of males found physics interesting, while the percentage was only 22% in the beginning.

## 2. Students' attitude towards learning methods

This section presents students' attitudinal results from interventions 3, 4 and 7. The figures given below present results in which there were weak gender effect observed and all students gave similar responses in all the three interventions.

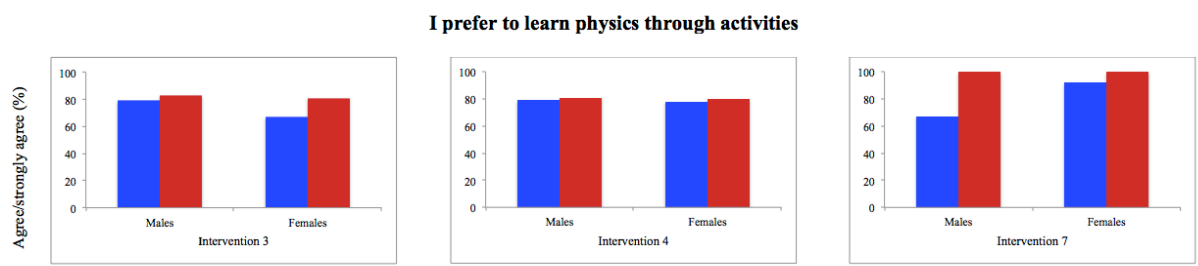

FIG. 4. Males and females response to "I prefer to learn physics through activities".

To assess the students' preference for learning method, we used the question "I prefer to learn physics through activities". In the Australian interventions (interventions 3 & 4), both genders preferred to learn physics through activities. Students' scores were already high in the pre-test and there was minimal increment in both genders after the programme. The improvement factor of boys and girls in intervention 3 is noted as 1.1 and 1.2, whereas in intervention 4, this factor is 1.2 and 1.0. However, after the programme in intervention 7, both genders agreed on the statement that learning through activities helped them to understand scientific ideas better than from just reading books. We also observed that the improvement factor for boys (1.5) is higher than for girls (1.1).

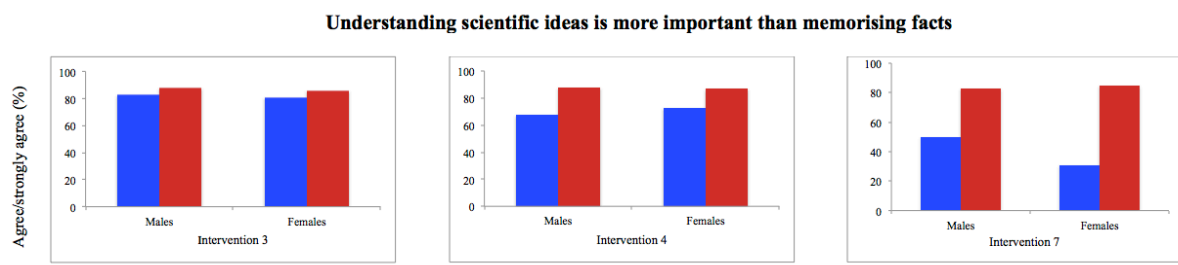

FIG. 5. Males and females response to "Understanding scientific ideas is more important than memorising facts".

As shown in figure 5, both genders also agreed with the statement that understanding new concepts were more important than rote learning. In all three interventions, students' attitude was generally positive even before the programme. In intervention 3, before the programme, 81% of boys and girls agreed with the statement and there was a slight change in students' responses after the programme. The improvement factor was calculated as 1.1 for both genders. However, in the case of intervention 4, 70% of girls and 68% of boys agreed in understanding concepts rather than memorising. The improvement factor was calculated as 1.3 for both genders. On the other hand, in Indonesia, initially, 50% of boys and 31% of girls believed that understanding is more important than memorising and after the programme, there was a dramatic change in students' responses. Males' percentage changed to 83% while females' percentage changed to 85%. The improvement factor was calculated as 1.7 for boys and 2.7 for girls. These results show that Australian students were aware that understanding any concept is important while Indonesian students did not agree with this statement before the programme. However after the programme, both Australian and Indonesian students came to know the importance of understanding any scientific idea as compared to memorising it.

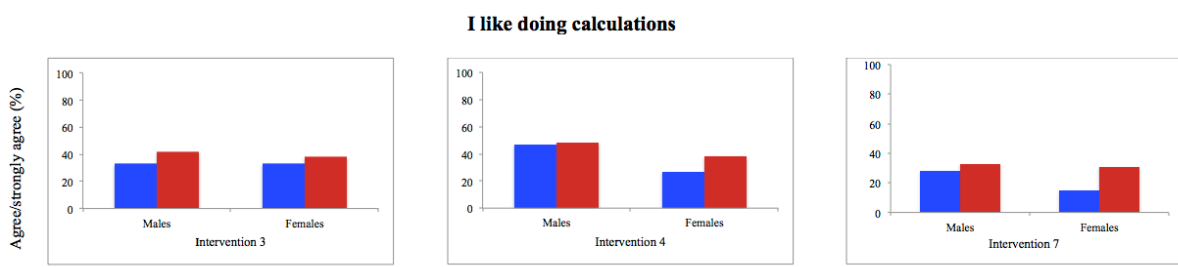

FIG. 6. Males and females response to "I like doing calculations".

To measure the students' interest in doing mathematical calculations, we asked students the question "I like doing calculations". In all interventions the scores are below 50% for all students as shown in figure 6. In both interventions in Australia, both genders did not like mathematical calculations. There was no any significant change observed after the programme. However, the percentage of girls in intervention 4 is slightly more as compared to their initial percentage. In Indonesia (intervention 7), initially, girls' (15%) interest was lower than boys (28%). After the programme, 30% of both genders like to do calculations. The improvement factor for girls is 2 and for boys it is 1.1.

Overall, both genders in both countries do not like mathematical calculations.

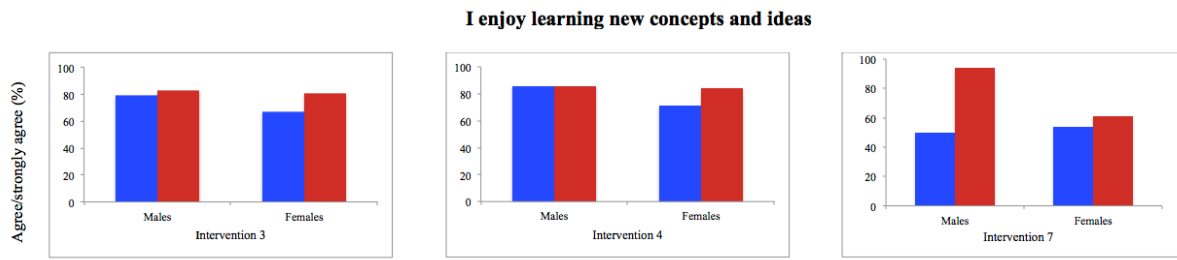

FIG. 7. Males and females response to "I enjoy learning new concepts and ideas". The figure shows that there is more female improvement in Australia, whereas males' improvement is greater in Indonesia.

Figure 7 show that both genders of Australian students (interventions 3 and 4) had very high pre and post scores, which indicate their enthusiasm for learning new concepts even before they knew about them. In both cases the improvement factor for girls was larger (1.2 in both interventions). In Indonesia (intervention 7) the cultural context could perhaps explain the lower scores, and in this case much greater receptivity to new concepts by the boys in the post-test. In Indonesia, the improvement factor for boys is 1.9 whereas for girls it is 1.1). The girls' improvement was positive in all cases, but in Indonesia, the improvement in boys was more dramatic.

## D. A brief discussion on knowledge and attitudinal results obtained from short interventions

Interventions 1, 2, 5, and 6 were similar but much shorter interventions across years 7-10 from the same school, aimed at discerning age dependent receptivity to Einsteinian concepts. Years 8, 9 and 10 were gifted and talented science students while year 7 were gifted and talented linguist students. The results of these interventions were extracted from another published paper [34]. The conceptual understanding scores are substantially lower than that of the programmes discussed above, indicating that short interventions are much less effective than longer ones. In spite of these differences, we observe a moderate rise in improvement for students of all age group. This implies that Einsteinian physics intervention is equally effective to everyone, regardless of a student's academic level or age group. Students' results of conceptual understanding are given below.

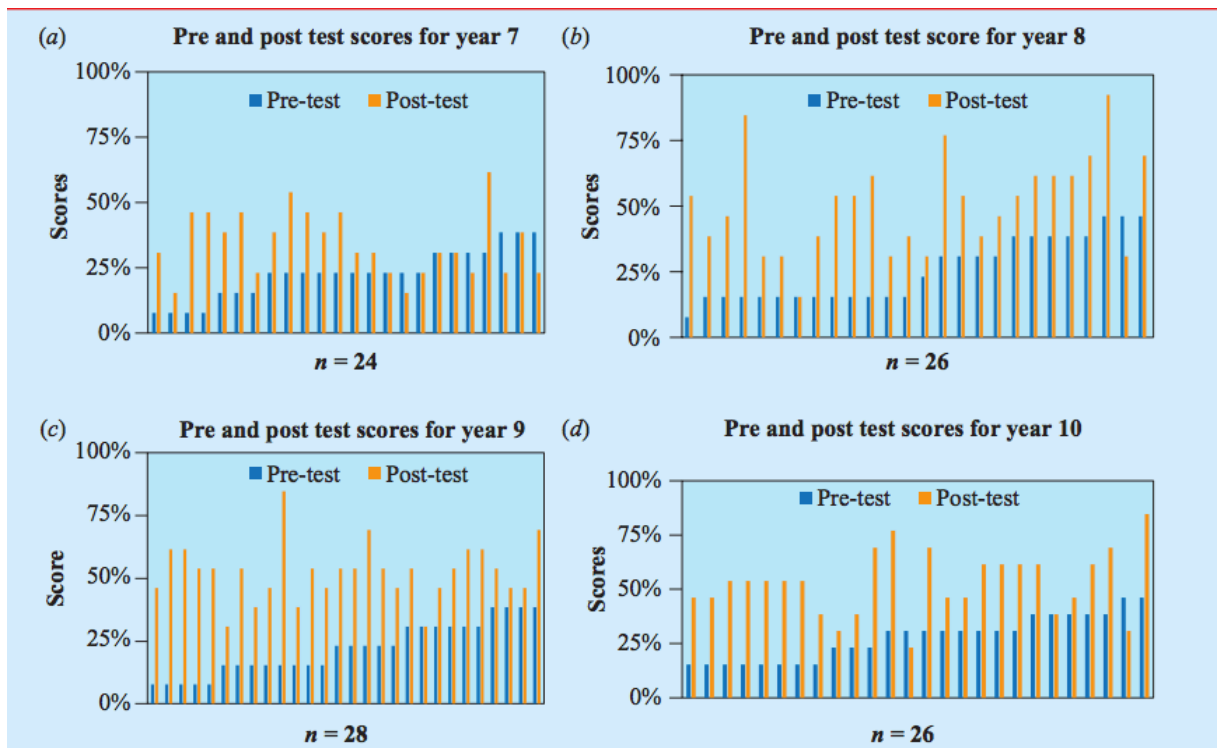

Fig. a) Pre and post scores for 24 students of Year 7 arranged in ascending order of the pre-test results. b) Pre and post scores for 26 students of Year 8 arranged in ascending order of the pre-test results. c) Pre and post scores for 28 students of Year 9 arranged in ascending order of the pre-test results. d) Pre and post scores for 26 students of Year 10 arranged in ascending order of the pre-test results.

Results from intervention 1: Figure 1a shows the results obtained from a one-day intervention with Years 7, 8, 9 and 10 (interventions 1, 2, 5 and 6 in Australia). It is clearly shown in the figure that after every programme, there is an improvement in students' scores. However, this improvement is low as compared to interventions 3, 4 and 7 (long interventions). The average scores of the students in interventions 1, 2 and 3 were noted as 37%, 50% and 50%. While this average was noted as 73% in intervention 3, 91% in intervention 4 and 78% in intervention 7.

On the other hand, if we compare girls' and boys' improvement, in interventions 1 – 7, the improvement factor for girls was found as 1.4, 2.4, 2.4, 4.3, 3.3, 1.9 and 3.3, while for boys, it was noted as 1.7, 1.5, 3.3, 3.8, 1.9, 2.0 and 3.7. These results clearly show that improvement factor for both gender is greater in longer interventions as compared to shorter interventions.

Overall, these results show that longer interventions are more effective compared to the shorter ones.

## Gender differences in attitude towards science according to short intervention

Students' attitude towards science is also observed in all four short interventions. The language of the questions asked in the short interventions were different from the long interventions, but the meaning of the questions was same. The results obtained from interventions 1, 2, 5 and 6 are given below.

A. Activity based learning: To assess students' attitude towards doing experiments, assessed the following two questions were asked:

  1) I would prefer to find out why something happens by doing an experiment than by being told.
  2) I would rather agree with other people than doing an experiment to find out for myself.

The results obtained from these two questions show that students entered the program with positive attitude towards learning by doing activities. There was not much room for improvement. Across all years, and both genders, students prefer to learn by doing experiments before and after the programme.

B. Science is only for smart people: This question was asked to test the stereotypical views that science is for nerds and smart people. A moderate improvement was observed in interventions 1, 2 and 6. While in intervention 5, improvement is girls' responses were higher with 1.7 improvement factor. The response to this question suggests that the program was able to demystify science among the students.

C. I would like to have a career in science: This question was asked to see whether students want to opt for a career in science. The results showed that the program had minimal effect on career choices. A small effect was observed among Years 8 and 9 boys. There was no effect observed among Year 7 and 10 students probably because they were selected language students.

D. Science classes teach me new things which are interesting: This question was asked to see students' interest in science classes. Students entered the program with high pre-scores. However, a small improvement was observed among Year 7 – 9 boys and Year 10 girls. The highest improvement was observed in Year 8 boys with 1.5 improvement factor.

E. Science classes would be more interesting if we learnt modern topics: This question was asked to see how much students were aware that the concepts they are learning in school are not modern concepts. We found that students were aware that the concepts they are learning in their school curriculum is not up to date. Across all years and both genders, students were interested to learn about modern topics. The maximum improvement was seen among Year 10 girls with 1.5 improvement factor.

In short interventions we observe that the improvement for both male and female is comparable. However this observation is not similar to the conceptual improvement, in which case the conceptual improvement is significantly higher for females compared to males. This observation is consistent with other longer interventions, which also indicate higher conceptual improvement of girls compared to boys.

When asked whether they would like to have a career in science, we observed that almost all the students quite universally agree before and after program. One probable reason for this observation could be that 77% of the total students were gifted and talented science students. Also if asked whether they would like to learn modern topics in classroom, the students' response change significantly and is higher for female students in three out of the four cases that we have observed. This observation provides us with a strong evidence of including modern concepts in high school science.

Overall, the above observations indicate that Einsteinian physics can be equally effective for both male and female students, thus providing us with strong evidence against the general interpretation where physics is believed to be more effective for males compared to female students.

## IV. DISCUSSION

Because the methods used for teaching Einsteinian physics involved all the methods that have already been shown to lead to improved female response to physics it is not surprising that this work has shown a strong gender effect. However, a-priori we had not reason to expect that conceptual learning of modern physics, which clearly is strongly embraced by both sexes, should have a positive gender effect. It could have been the opposite.

However, results from the question on "liking to learn new concepts" shows a weak positive gender effect for Australian students, and all results show an increased score on this question. (We note that the very high pre-test scores on this question (most students scoring agree or strongly agree) makes the resolution of changes rather weak in a Likert scoring system.)

Thus, we cannot claim that conceptual learning of modern physics is the cause of the gender equalising effects. However, the necessity of using activity based learning with group learning based on models and analogies has created a learning environment that has a gender equalising effect.

Data from all the interventions make it clear that most students have some conception that Einsteinian science is important. We suggest that this is probably because Einsteinian concepts such as black holes, light speed and time warps, as well as people like Einstein and Stephen Hawking, are often depicted in the popular media such as The Simpsons, Big Bang Theory, Star Wars and Doctor Who. Whether or not students understand the nature of Einsteinian physics, they are certainly aware that these are topics rarely encountered in school. Thus, there is a discrepancy between science as frequently portrayed and the science encountered in class.

We considered whether the sex of the presenter was relevant. Our interventions have had more male presenters than females (two females, five males), but similar results have been obtained in all interventions, and in particular comparison of interventions 4 (female) and 7 (male) indicate that the sex of the presenter is not a strong factor.

Noting that boys have an initially higher attitude score than girls, the observed gender-equalizing effect of the Einsteinian physics intervention could be because latent gender bias in favour of boys is absent in the Einsteinian physics interventions. Some educationalists claim that teaching about projectiles in Newtonian physics represents such a bias, because girls are less interested in weapons. However for teaching the key concept of photons we always use Nerf guns in which the Nerf gun bullets are used as analog photons. Clearly the suggested male bias towards use of weapons has not been a significant factor here.

We ask whether girls intrinsically respond more strongly to conceptual learning than boys. We are introducing questions such as what is space, what is time and what is light. These questions are rarely addressed in school. The Newtonian and Euclidean concepts of space and time are implicit in graphs, gridlines and assumptions about time being absolute. We address the fundamentals by explicitly teaching the concepts of curved space and time depending on altitude, and light being a stream of photons.

Our repeated finding that knowledge scores for simple knowledge questions [9] are uncorrelated with pre-test results shows that it is neither pre-exposure nor academic talent that determines student ability to grasp the concepts of Einsteinian physics.

We suggest that the positive gender effect we report here is due primarily to the gender bias that comes from accumulated prior experience of science, as a result of conventional teaching which does not have a sufficient level of interactivity, group learning and perceived relevance. The gender neutrality of the programmes we present, combined with the fact that the concepts are completely new, allows males and females to attain similar scores after the intervention.

From all different interventions we can draw the following conclusions:

- There is significant improvement in students understanding after every programme, but the magnitude of the improvement depends on the duration. Single day interventions create strong and statistically significant improvements in student knowledge and attitude, but multi-week interventions are much more effective
- There is significant improvement in student's attitudes toward science in all interventions. The female improvement factor is generally most pronounced.
- At the end of every programme, it was found that female students have  higher  attitude scores when compared to those of the male students' scores.

## V. CONCLUSION

We have presented evidence that female students respond more positively than male students to interventions that seek to introduce students to Einsteinian science. Across a range of interventions from a single day to 20 lessons over 10 weeks, and over an age range from 11 to 15, a positive gender effect has been reproduced.

The positive is likely to be due to the interactive methods that we found necessary to introduce the fundamental concepts of Einsteinian physics to young people, combined with students perception of the perceived relevance of Einsteinian physics, and a negative perception of the relevance of their conventional physics.

We believe that the gender effect size reported here and other positive outcomes reported elsewhere [35] is sufficiently strong to justify revision of school curricula to include the Einsteinian understanding of the world around us. Given that common sense "is the prejudices acquired by the age of 18" [36], we predict that future generations who have learnt Einsteinian science at an early age will accept quantum interference as common sense, and curved space and time dilation as self evident. Most importantly, if taught by the methods we have demonstrated, all student attitudes to science should improve, and the gender gap in attitude to science should be greatly diminished.

## References

[1]  C. Andrew ed., Girls and Science: A Training Module on Motivating Girls to Embark on Science and Technology Careers, report from Division of Secondary, Technical, and Vocational Education, New York: UNESCO, 34. (2007).
[2]  A. Ekine,  https://www.brookings.edu/wp-content/uploads/2016/07/ekine_girls_education.pdf. (2016).
[3]  Economic_benefits_of_gender_equality_briefing_paper.pdf
[4]  E. Unterhalter et al., Interventions to enhance girls' education and gender equality, Education Rigorous

Literature Review, Department for International Development (2014).

[5] E. Gillibrand, P. Robinson, R. Brawn, and A. Osborn, Girls' participation in physics in single sex classes in mixed schools in relation to confidence and achievement, International Journal of Science Education 21, 349 (1999).

[6] K. Tolley, The Science Education of American Girls: A Historical Perspective (RoutledgeFalmer, New York, (2003).

[7] D. F. Halpern, J. Aronson, N. Reimer, S. Simpkins, J. R. Star, and K. Wentzel, Encouraging Girls in Math and Science (National Center for Education Research, Washington, DC, (2007).

[8] T.A. Huebner, Encouraging girls to pursue math science, Educ. Leader 67, 90 (2009).

[9] A. M. Kelly, Social cognitive perspective of gender disparities in undergraduate physics, Physical Review Physics Education Research 12, 020116, 2016

[10] R. Koul, T. Lerdpornkulrat, and C. Poondej, Gender compatibility, math-gender stereotypes, and self-concepts in math and physics, Physical Review Physics Education Research 12, 020115, 2016.

[11] J. M. Nissen and J. T. Shemwell, Gender, experience, and self-efficacy in introductory physics, Physical Review Physics Education Research 12, 020105, 2016.

[12] E. A. Dare and G. H. Roehrig, "If I had to do it, then I would": Understanding early middle school students' perceptions of physics and physics-related careers by gender, Physical Review Physics Education Research 12, 020117 (2016).

[13] National Academies, Beyond Barriers: Fulfilling the Potential of Women in Academic Science and Engineering (National Academies Press, Washington, DC, (2007).

[14] J. Sikora and A. Pokropek, Gender segregation of adolescent science career plans in 50 countries, Science Education 96, 234 (2012).

[15] R. H. Tai, C. Q. Liu, A. V. Maltese, and X. Fan, Planning early for careers in science, Science 312, 1143 (2006).

[16] S. Brophy, S. Klein, M. Portsmore, and C. Rogers, Advancing engineering education in P-12 classrooms, J. Eng. Educ. 97, 369 (2008).

[17] A. Calabrese Barton, E. Tan, and A. Rivet, Creating hybrid spaces for engaging school science among urban middle school girls, Am. Educ. Res. J. 45, 68 (2008).

[18] L. Archer, J. Dewitt, J. Osborne, J. Dillon, B. Willis, and B. Wong, "Doing" science versus "being" a scientist: Examining 10/11-year-old schoolchildren's construction of science through the lens of identity, Sci. Educ. 94, 617 (2010).

[19] R. H. Tai, C. Q. Liu, A. V. Maltese, and X. Fan, Planning early for careers in science, Science 312, 1143 (2006).

[20] P. Häussler and L. Hoffmann, A curricular frame for physics education: Development, comparison with students' interests, and impact on students' achievement and self-concept, Sci. Educ. 84, 689 (2000).

[21] P. Labudde, W. Herzog, M. P. Neuenschander, E. Violi, and C. Gerber, Girls and physics: Teaching and learning strategies tested by classroom interventions in grade 11, Int. J. Sci. Educ. 22, 143 (2000).

[22] P. Häussler and L. Hoffmann, A curricular frame for physics education: Development, comparison with students' interests, and impact on students' achievement and self-concept, Sci. Educ. 84, 689 (2000).

[23] P. Häussler and L. Hoffmann, An intervention study to enhance girls' interest, self-concept, and achievement in physics classes, J. Res. Sci. Teach. 39, 870 (2002).

[24] K.Wilson, D. Low, M. Verdon, and A. Verdon, Differences in gender performance on competitive physics selection tests, Physical Review Physics Education Research 12, 020111 (2016).

[25] P. Labudde, W. Herzog, M. P. Neuenschander, E. Violi, and C. Gerber, Girls and physics: Teaching and learning strategies tested by classroom interventions in grade 11, Int. J. Sci. Educ. 22, 143 (2000).

[26] I. Rodriguez, G. Potvin, and L. H. Kramer, How gender and reformed introductory physics impacts student success in advanced physics courses and continuation in the physics majore, Physical Review Physics Education Research 12, 020118 (2016).

[27] E. A. Dare and G. H. Roehrig, "If I had to do it, then I would": Understanding early middle school

students' perceptions of physics and physics-related careers by gender, Physical Review Physics Education Research 12, 020117 (2016).

[28]  T. Kaur, D. Blair, J. Moschilla, W. Stannard, and M. Zadnik, Teaching Einsteinian physics at schools: part 1, models and analogies for relativity, Physics Education **52**, 065012 (2017).

[29] P. R. Saulson, Josh Goldberg and the physical reality of gravitational waves, General Relativity and Gravitation **43**, 3289 (2011).

[30] R. Feynman, QED: The Strange Theory of Light and Matter, Alix G. Mautner Memorial Lectures, New Jersey: Princeton University Press, p.15 (1985).

[31] T. Kaur, D. Blair, J. Moschilla, W. Stannard, and M. Zadnik, Teaching Einsteinian physics at schools: part 1, models and analogies for relativity, Physics Education **52**, 065012 (2017).

[32] T. Kaur, D. Blair, J. Moschilla, W. Stannard M. Zadnik, Teaching Einsteinian physics at schools: part 2, models and analogies for quantum physics, Physics Education **52,** 065013 (2017).

[33] T. Kaur, D. Blair, J. Moschilla, W, Stannard and M. Zadnik, Teaching Einsteinian physics at schools: part 3, Review of research outcomes, Physics Education **52,** 065014 (2017).

[34] R. Kumar, A. Foppoli, T. Kaur, D. Blair, M. Zadnik and R. Meagher, Can a short intervention focused on gravitational waves and quantum physics improves students' understanding and attitude?, Physics Education 53, 06 (2018)

[35] T. Kaur, D. Blair, R. Burman, W. Stannard, D. Treagust, G. Venville, M. Zadnik, W. Mathews, and D. Perks, Evaluation of 14 to 15 Year Old Students' understanding and attitude towards learning Einsteinian physics, https://arxiv.org/abs/1712.02063, (2017).

[36] A. Calaprice, The Ultimate Quotable Einstein, Section: Probably Not By Einstein, Page 481, Princeton University Press, Princeton, New Jersey, (2010).